\documentclass[aps,prb,twocolumn,superscriptaddress]{revtex4-2}
\usepackage{amsfonts}
\usepackage{graphicx}
\usepackage{epstopdf}
\usepackage{epsfig}
\usepackage{amsmath}
\usepackage[normalem]{ulem}
\usepackage{multirow}
\usepackage{color}
\usepackage{makecell}
\usepackage{array}
\usepackage{booktabs}
\usepackage{mathtools}
\usepackage{bm}
\usepackage{color}
\usepackage{array}
\usepackage{booktabs}
\usepackage{tabularx}

\newcolumntype{Y}{>{\centering\arraybackslash}X}

\newcommand{\Rmnum}[1]{\expandafter\@slowromancap\romannumeral #1@}
\makeatother

\begin{document}

\title{Generating extreme quantum scattering in graphene with machine learning}

\author{Chen-Di Han}
\affiliation{School of Electrical, Computer and Energy Engineering, Arizona State University, Tempe, Arizona 85287, USA}

\author{Ying-Cheng Lai} \email{Ying-Cheng.Lai@asu.edu}
\affiliation{School of Electrical, Computer and Energy Engineering, Arizona State University, Tempe, Arizona 85287, USA}
\affiliation{Department of Physics, Arizona State University, Tempe, Arizona 85287, USA}

\begin{abstract}

	Graphene quantum dots provide a platform for manipulating electron behaviors in two-dimensional (2D) Dirac materials. Most previous works were of the ``forward'' type in that the objective was to solve various confinement, transport and scattering problems with given structures that can be generated by, e.g., applying an external electrical field. There are applications such as cloaking or superscattering where the challenging problem of inverse design needs to be solved: finding a quantum-dot structure according to certain desired functional characteristics. A brute-force search of the system configuration based directly on the solutions of the Dirac equation is computational infeasible. We articulate a machine-learning approach to addressing the inverse-design problem where artificial neural networks subject to physical constraints are exploited to replace the rigorous Dirac equation solver. In particular, we focus on the problem of designing a quantum dot structure to generate both cloaking and superscattering in terms of the scattering efficiency as a function of the energy. We construct a physical loss function that enables accurate prediction of the scattering characteristics. We demonstrate that, in the regime of Klein tunneling, the scattering efficiency can be designed to vary over two orders of magnitudes, allowing any scattering curve to be generated from a proper combination of the gate potentials. Our physics-based machine-learning approach can be a powerful design tool for 2D Dirac material-based electronics.

\end{abstract}

\date{\today}

\maketitle

\section{Introduction} \label{sec:intro}

Two-dimensional (2D) Dirac material systems such as graphene, topological
insulators, molybdenum disulfied, and topological Dirac
semimetals~\cite{novoselov2004electric,novoselov2005two,neto2009electronic,wehling2014dirac,wang2015rare} constitute a research frontier in condensed
matter physics and materials science. A common feature of these materials is
that their energy bands contain a Dirac cone structure that gives rise to
a linear energy-momentum relation (dispersion relation) for low-energy
excitations, which is characteristic of relativistic quantum motions governed
by the Dirac equation. The Dirac cone structure can be exploited for storing
and transferring information with applications in Dirac-material based
electronics and spintronics~\cite{han2014graphene,KCSVKDRD:2018,Kirmani:2019,AOGOvWV:2020}.

Given a 2D Dirac material, the system structure and the applied gate
potentials, a combination of these can generate rich quantum behaviors.
For example, in the field of relativistic quantum chaos~\cite{huang2009relativistic,xu2013chiral,huang2018relativistic,lai2018relativistic},
electrons can be confined in a graphene sheet of certain geometric shape,
generating billiard systems that provide a platform to study the relativistic
quantum manifestations of different kinds of classical dynamics. In scattering,
a combination of gate potentials can be applied to a graphene sheet to create
a quantum-dot structure, which can be experimentally realized through STM tips
or doping~\cite{zhao2015creating,lee2016imaging,gutierrez2016klein,velasco2016nanoscale,ghahari2017off,ge2021imaging}.
Graphene quantum dots are also a paradigm to study various transport
phenomena~\cite{brey2006electronic,son2006energy,rycerz2007valley}. Given a
scattering structure, the tight-binding Hamiltonian or the Dirac equation can
be solved to yield the scattering functions so that the performance of the
device can be assessed. A common feature among all these previous works is
their ``forward'' nature: studying various quantum behaviors or performance
of the underlying system with a {\em given} structure.

Inverse design addresses the opposite problem: how to design a Dirac material
system to generate certain desired functional characteristics. For example,
suppose we wish to design a graphene device to generate the desired scattering
functions by using a 2D multilayer structure of concentric circular graphene
ribbons, where the layers can have different widths and each layer is made
physically distinct from the others through a vertical electric field (gate
potential). A brute-force approach to searching the optimal multilayer structure
to generate the desired scattering functions would be to test a large number of
combinations of the geometric parameters of the various layers as well as the
values of the gate potentials. With luck, it may be possible to find a specific
structure and a set of gate potentials such that the resulting scattering
curves approximately match the desired functions. However, such a brute-force
approach is generally practically infeasible because of the extremely large
parameter space that needs to be searched. There is in fact no guarantee that
this approach would be successful, as the existence of a device with the
desired response is unknown a priori. Besides quantum
scattering~\cite{apagyi1996inverse}, inverse-design problems of this kind
occur in other fields such as quantum information~\cite{banchi2016quantum},
biology~\cite{ma2009defining}, molecular design~\cite{sanchez2018inverse}, 
and photonics~\cite{molesky2018inverse}.

Recently, machine learning has been introduced to inverse optical design where
the goal is to find the best structure of a multilayer dielectric sphere to
generate the desired electromagnetic response by approximating the Maxwell
equations with a trained artificial neural
network~\cite{peurifoy2018nanophotonic}. The approximation can simplify the
original optimization problem without loss of details. Subsequently, the
approach has been extended to designing optical
metasurfaces~\cite{liu2018generative}, metagrating~\cite{jiang2019free}, and
multifunctional devices~\cite{an2021multifunctional,zhu2021building}. Machine
learning has thus offered a general solution to the inverse-design problem.

This work concerns designing systems that can generate {\em extreme} scattering
characteristics. In particular, the performance of any scattering system,
optical or electronic, can be conveniently characterized by the scattering
cross section as a function of a basic quantity such as the wavelength (in
optics) or the Fermi energy (in electronic systems). The two opposite extremes
correspond to a small and a large scattering cross section, respectively,
with the former representing cloaking and the latter signifying
superscattering. In optics, one approach to cloaking is through scattering
cancellation~\cite{alu2005achieving} based on the the idea that, in a
multilayer structure, polarization from different layers can cancel each
other to generate an exceedingly small scattering cross section.
Multilayer structures can also be exploited to produce
superscattering~\cite{qian2019experimental}. There were experimental results
on cloaking or superscattering in optics~\cite{schurig2006metamaterial,edwards2009experimental,qian2019experimental}.
For those problems, a basic physical constraint is that it is generally not
possible to generate cloaking or superscattering for all kinds of incident
waves~\cite{monticone2013cloaked,fleury2015invisibility}, so these exotic
phenomena can occur only for a certain type of incident waves.
In optics, both cloaking and superscattering can occur in a multilayer
structure of dielectric materials, so it is possible to generate cloaking and
superscattering in the same device. For example, it has been recently
demonstrated that a single device based on Ag-Semiconductor multilayer
sphere can produce large and small scattering cross
sections~\cite{peurifoy2018nanophotonic, sheverdin2020photonic}. Without
changing the dimension of the device, a superscattering-cloaking transition
was reported in a two-layer system where a significant change in the
scattering cross section can occur with respect to the incident
angle~\cite{verslegers2012electromagnetically}. One optical material capable of
such a transition is GeTe~\cite{huang2018switching,lepeshov2019nonscattering},
where, under different environmental conditions such as the
temperature~\cite{luo2021deep}, the material exhibits different physical
properties. Related optical scattering problems were investigated with
graphene coated nanosphere~\cite{farhat20133d,li2015atomically} and its
design~\cite{li2016design}. The analogy between matter and electromagnetic
waves stimulated the idea of generating electron cloaking through a multilayer
structure~\cite{zhang2008cloaking,liao2012cloaking,fleury2013quantum},
which subsequently was extended to graphene for cloaking relativistic
electrons~\cite{liao2013isotropic,oliver2015cloaking,sadrara2022collective}.

Our study focuses on the following question: is it possible to generate the
two extreme scattering behaviors, cloaking and superscattering, in the same
2D Dirac material system? Because of the relative easiness to experimentally
tune the gate potential through varying an applied electrical
field~\cite{chen2011atomically}, graphene stands out as a viable candidate
for generate superscattering and cloaking as well as a transition between
them in the same fixed structure. To be concrete, we consider a multilayer
scattering structure that consists of $N$ concentric circles, where a different
gate potential is applied to each distinct circular layer. To reduce the
dimension of the parameter space so as to make the inverse-design problem
feasible, we fix the geometric structure of the scattering system and allow
the set of gate potentials to be tunable. We develop a class of neural networks
subject to physical constraints with the goal to find a set of gate potentials
to generate electronic cloaking and another set to generate superscattering in
the {\em same device}. A key component of the inverse design is our
articulation of a physical loss function to significantly reduce the training
and testing errors and to eliminate nonphysical solutions. We demonstrate that,
even in the regime of Klein tunneling, the scattering efficiency (to be defined
in Sec.~\ref{subsec:GQD}) can vary over two orders of magnitude. Compared with
the corresponding optical system where the material for each layer is fixed
and the width for each shell can be changed, in our graphene scattering system
only the gate potential is changed with the device structure intact, rendering
it experimentally feasible. Our physics-based machine-learning approach can be
a powerful design tool for graphene-based electronic devices and is
generalizable to solving inverse problems in other areas of science.


\section{Methods: multilayer graphene quantum dot, machine learning and inverse design} \label{sec:methods}

\begin{figure} [ht!]
\centering
\includegraphics[width=\linewidth]{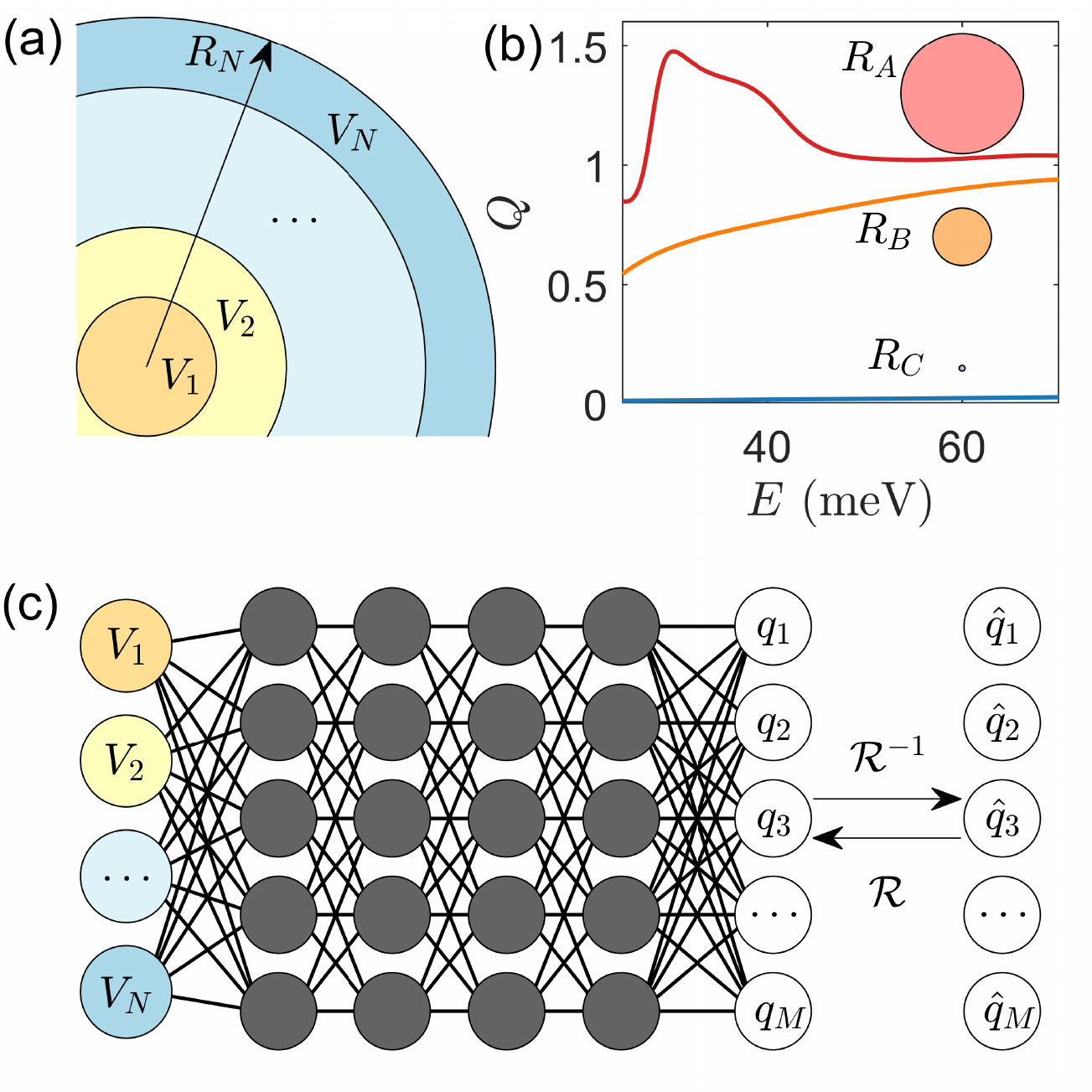}
\caption{Illustration of a multilayer graphene quantum dot, scattering
efficiency, and the physics-constrained machine learning scheme.
(a) A multilayer circular graphene quantum dot generated by the gate potential
profile $V(r)$ in Eq.~(\ref{eq:potential_profile}). (b) For the special case of
a single-layer quantum dot, scattering efficiency versus the Fermi energy $E$
for three different sizes of the quantum dot: $R_A = 63$ nm, $R_B = 30$ nm, and
$R_c = 3$ nm. Because of the simple structure, the overall scattering
efficiency increases as the dot becomes larger, as expected, indicating
naturally that a single-layer structure is not capable of generating extreme
scattering behaviors. (c) The neural-network architecture underlying our
proposed scheme of physics-constrained machine learning. The basic component
of the neural network is a multilayer perceptron, where the input is the set
of constant gate potential values $(V_1,V_2,\ldots,V_N)$ and the output is the
properly discretized scattering efficiency function as exemplified in (b). The
function is defined in a fixed energy range and is uniformly sampled with $M$
discrete points, leading to an $M$-dimensional output vector:
$\bm{Q} \equiv (q_1,q_2,\ldots,q_M)^T$. The rightmost layer with the output
vector $\bm{\hat{Q}} \equiv (\hat{q}_1,\hat{q}_2,\ldots,\hat{q}_M)^T$ is
introduced to ensure that the whole neural-network architecture ``respects''
the basic physics of quantum scattering (see text for details).}
\label{fig:schematic}
\end{figure}

We first clarify the physical meanings of the terms ``cloaking'' and
``superscattering'' adopted in this work. Our goal of inverse design is to
realize a structure that can generate these two opposite extremes of quantum
scattering. Ideal cloaking is characterized by zero scattering cross section
or efficiency. Since near-zero scattering efficiency can trivially occur for a
scatterer of arbitrarily weak strength: $\bar{V}\bar{R} \rightarrow 0$, where
$\bar{V}$ and $\bar{R}$ are the average gate potential and the mean size of
the quantum dot, respectively, cloaking is meaningful only for scatterers with
a reasonably large scattering strength, e.g., the quantum dot of size $R_B$ in
Fig.~\ref{fig:schematic}(b). For this case to be qualified as cloaking, the
resulting scattering cross section should be as small as that from a scatterer
of much smaller size such as the dot of radius $R_C$ in
Fig.~\ref{fig:schematic}(b). In the opposite limit, strong scattering can
naturally occur if the scattering strength is high. By superscattering we mean
that, even when the scatterer is relatively weak, the resulting scattering
cross section can be as large as that from a scatterer of much larger size,
such as the quantum dot of size $R_A$ in Fig.~\ref{fig:schematic}(b).

\subsection{Graphene quantum dot} \label{subsec:GQD}

We consider a 2D circular graphene quantum dot of radius $R_N$ subject to a
gate potential profile $V(\mathbf{r})$. The Hamiltonian of the scattering
system is
\begin{align} \label{eq:Hamiltonian}
H=v_g\bm{\sigma}\cdot \mathbf{p}+V(\mathbf{r}),
\end{align}
where $v_g$ is the Fermi velocity and $\bm{\sigma} \equiv [\sigma_x,\sigma_y]$
is the vector of Pauli matrices. For simplicity, we assume that the gate
potential has no angular dependence and write $V(\mathbf{r}) = V(r)$, for
$0 < r \le R_N$. We further assume that $V(r)$ is a piecewise constant
function with $N$ distinct values, which experimentally can be realized by
STM tips or doping in the substrate~\cite{zhao2015creating,lee2016imaging,gutierrez2016klein,velasco2016nanoscale,ghahari2017off,ge2021imaging}. The potential
profile is
\begin{align} \label{eq:potential_profile}
V(r)=\begin{dcases}
V_1, & r<R_1 \\
V_2, & R_1<r<R_2 \\
	\cdots & \cdots \\
V_N, & R_{N-1}<r<R_N \\
0, & r>R_N
\end{dcases},
\end{align}
leading to a multilayer scattering structure with $N$ distinct concentric
circular layers, where the gate potential in each layer is a constant, as
schematically illustrated in Fig.~\ref{fig:schematic}(a). An advantage of the
multilayer structure is that the structural parameters
$(\mathbf{R},\mathbf{V})\equiv([R_1,R_2,\cdots,R_N]^T,[V_1,V_2,\cdots, V_N]^T)$
can be relatively readily adjusted in experiments. This should be contrasted
to a multilayer structure in optics~\cite{peurifoy2018nanophotonic,liu2018generative,jiang2019free,an2021multifunctional,zhu2021building}, where only the
set of radii can be changed once the dielectric materials are fixed.

Our goal is to design the potential profile $V(r)$ to generate extreme
scattering behaviors for a fixed set of radii $(R_1,R_2,\ldots,R_N)$. To gain
insights, we first consider the relatively simple case of a single-layer
scatterer: $N = 1$. Matching the boundary condition at $r=R_1$ and setting the
Fermi velocity to be $v_g=10^6$ m/s, we obtain the solutions of the 2D Dirac
equation~\cite{novikov2007elastic} in terms of the scattering spinor
wavefunction (Appendix~\ref{Appendix_A}). The scattering behavior can be
conveniently characterized by the scattering efficiency defined as
\begin{align} \label{eq:scattering_efficiency}
Q=\frac{\text{Scattering Cross Section}}{\text{Geometric Size}}=\frac{2}{kR_1}\sum_{l=-\infty}^{\infty} |A_l|^2,
\end{align}
where $A_l$ is the coefficient for scattering wave in the polar coordinates
associated with angular momentum $l$. Intuitively, a large scattering
efficiency indicates that the object is more ``visible.'' Three examples of
the scattering efficiency are shown in Fig.~\ref{fig:schematic}(b) for
$V_1 = 87.5$ meV and for dot size $R_A = 63$ nm, $R_B = 30$ nm, and $R_C = 3$
nm, corresponding to strong, intermediate, and weak (all in the relative sense)
scattering, respectively. Note that the $Q$ value for $R_B$ is approximately
two orders of magnitude larger than that for $R_C$. 
Note also that the incident energy is approximately one-quarter to 
three-quarter of the potential height, so the particle is in the Klein 
tunneling regime~\cite{KNG:2006}.
A recent work~\cite{wang2021super} indicated that, for a graphene quantum dot, 
in the limit $(kR_1)\rightarrow 0$, for a fixed Fermi energy the scattering 
efficiency scales with the size of the scatterer as $Q\propto (VR_1)^3$. 
However, the scaling breaks down for $R_1\approx 10$ nm at which the scattering
efficiency tends to saturate and increases only slowly with further increase 
in $R$. Overall, Fig.~\ref{fig:schematic}(b) indicates that, as the size of the
scatterer increases, scattering is enhanced, as can be intuitively anticipated.
The simple single-layer structure is thus not capable of generating extreme
scattering behaviors.

For a graphene multilayer circular quantum dot with given parameters
$(\mathbf{R},\mathbf{V})$, the scattering problem can be solved by the method
of transfer matrix~\cite{nguyen2016transfer} through matching the spinor
wavefunctions at all the layer boundaries (Appendix~\ref{Appendix_A}). After
the spinor wavefunctions in all layers have been obtained, we can calculate
the scattering efficiency according to
\begin{equation} \label{eq:2_Q}
Q=\frac{2}{k_{N+1}R_N}\sum_{l=-\infty}^{\infty} \left|A_l^{N+1}\right|^2,
\end{equation}
where $k_N \equiv |E|/v_g$. Given a scattering structure, it is thus
straightforward to calculate the scattering efficiency (the solutions of the
forward problem).

To generate extreme scattering behaviors, i.e., cloaking or superscattering,
it is necessary to define the input properly by specifying a range for the
gate potentials because, if the potentials are too weak, the scattering
efficiency will be trivially near zero and this does not lead to cloaking.
Likewise, unreasonably large potential values can lead to strong scattering,
but this is not superscattering. Making use of the scaling relation for
the scattering efficiency for weak scatters~\cite{wang2021super}:
\begin{align} \nonumber
Q \propto \left[\sum{|V_i|(R_{i+1}-R_i)}\right]^3,
\end{align}
we have $Q \propto (\langle |\mathbf{V}|\rangle R_B)^3$ because $R_{i}=i R_N/N$.
It is thus reasonable to choose the gate potentials such that
$\langle |\mathbf{V}|\rangle\approx V_0$, where $\langle |\mathbf{V}|\rangle$
is the average gate potential applied to the multilayer structure and $V_0$
is the potential applied to a single-layer structure of the same size.

For a single-layer scatterer, the problem of finding a scattering configuration
to generate the desired scattering efficiency $Q_\text{desired}$ can be
formulated as the following optimization problem
\begin{equation}
	\underset{\mathbf{V}}{\min}  \text{ }\left\| \frac{Q(\mathbf{V})-Q_\text{desired}}{Q_\text{desired}} \right\|^2,
\end{equation}
where $\| \cdot\|^2$ denotes the mean-square error (MSE).
Because of the simple geometry, it is only necessary to optimize a small
number of parameters, making it relatively straightforward to obtain the
solutions of the optimization problem. In particular, it is only necessary
to discretize each parameter dimension and perform a grid search. Difficulties
arise when the scatterer has many layers, making the function $Q(\mathbf{V})$
highly nonlinear with many parameters that need to be optimized. For example,
for a multilayer sphere of $10$ layers, even if only four parameters need to
be determined for each layer, the number of parameter combinations will be
larger than $4^{10} \sim 10^6$. This issue of computational complexity
motivated us to exploit machine learning to solve the optimization problem.

\subsection{Physics-constrained machine learning} \label{subsec:PCML}

Machine learning provides an effective platform to approximate mathematical
functions arising from physics~\cite{cybenko1989approximation}. For example,
in photonic design, neural networks can be used to approximately represent
the scattering cross section from a variety of
devices~\cite{peurifoy2018nanophotonic,liu2018generative}. To exploit machine
learning to generate extreme scattering behaviors in a graphene quantum dot,
an essential step is to train a multilayer perceptron such that it mimics the
scattering process. To achieve this, we introduce a physics-constrained neural
network architecture, as illustrated in Fig.~\ref{fig:schematic}(c). The
concrete design of the architecture is as follows. The input vector has the
dimension $N$ - a set of constant gate potential values
$\bm{V} \equiv (V_1,V_2,\ldots,V_N)^T$. The output is the scattering efficiency
function as exemplified in Fig.~\ref{fig:schematic}(b), which is defined in a
given energy range and is uniformly sampled with $M$ discrete points, leading
to an $M$-dimensional output vector: $\bm{Q} \equiv (q_1,q_2,\ldots,q_M)^T$.
In this work, we use $M = 200$ (a rather arbitrary choice). Besides the input
and output layers, the neural network has four hidden layers, each containing
$200$ neurons. There are thus five layer-to-layer transforms from the input to
the output. The activation for the first four transforms is set to be the
standard Rectified Linear Unit (ReLU) and there is no activation from the
fourth hidden layer to the output layer (a simple linear transform). Note that,
before any activation, the neural network uses a linear transformation between
two adjacent layers, which has approximately $4\times 10^4$ parameters. Thus
the number of training parameters in the architecture from the input vector
$\bm{V}$ to the output vector $\bm{Q}$ is approximately $10^5$.

To train the neural network, a certain amount of ground truth data from the
graphene scattering system is required. To generate the data, it is necessary
to determine the range of the input - which in our case is the range of
the externally applied gate potential. To be concrete, we assume that the
mean value of the gate potential in the whole quantum-dot scattering region
is $\langle V(r) \rangle \approx V_0$, where $V_0 = 87.5$ meV, so we choose
$|V_i| \in (75 \text{ meV}, 100 \text{ meV})$. The training data are generated
by randomly select $V_i$ in this range, where $V_i$ can be either positive or
negative, i.e., $-100\ \text{meV}< V_i<-75\ \text{meV}$ or
$75\ \text{meV}< V_i<100\ \text{meV}$. For a larger number $N$ of layers, the
multilayer quantum-dot structure is geometrically more sophisticated, thereby
requiring more training data. Empirically, we choose the number of training
data points to be $10^3N$. Say the scattering structure has 10 circular layers:
$N = 10$. Compared with a brute-force grid search in 10 dimensions, our choice
of $10^4$ data points corresponds to search in a parameter space of a
significantly reduced dimension: about three. Our machine-learning approach
thus requires far less data amount than a brute-force search would.

The training process is rather standard and is briefly described, as follows.
The training datasets are generated by directly simulating quantum scattering
based on the solutions of the Dirac equation in the setting of a multilayer
quantum dot. Training is conducted by employing the standard stochastic
gradient descent algorithm with batch size $128$. The method of adaptive
momentum (Adam) is used to minimize the loss function to find the parameters.
The whole process from network construction and training to predicting the
scattering-efficiency function is accomplished by using the open source
package Tensorflow and Keras~\cite{abadi2016tensorflow,chollet2015keras}.

\begin{figure} [ht!]
\centering
\includegraphics[width=\linewidth]{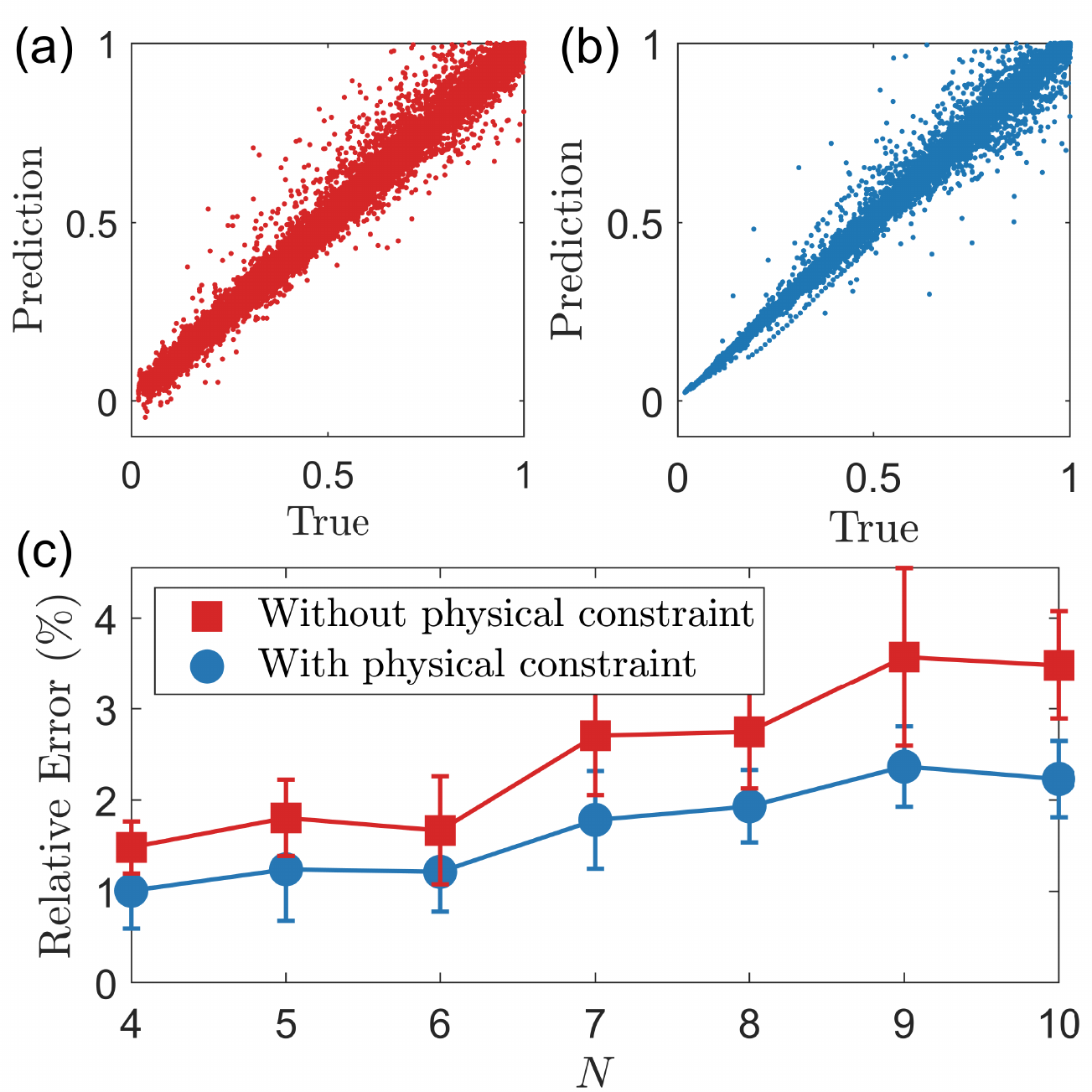}
\caption{The advantages of invoking a physics-constrained loss function. The
performance of a given loss function can be conveniently visualized by the
plot of the neural-network predicted values of the scattering efficiency
function versus the true value. (a) The plot based on the mathematical loss
function Eq.~\eqref{eq:error_MSE} (without physical considerations). There is
a relatively wide spread of the predicted values about the true values and
some predicted values of the scattering efficiency are negative (non-physical).
(b) With the physics-constrained loss function Eq.~\eqref{eq:error_phy}, the
predicted values are all positive and are closer to the true values.
(c) Relative prediction error versus the number of layers in the scattering
structure. Compared with the case of loss function Eq.~\eqref{eq:error_MSE},
the relative error associated with the physics-constrained loss function
Eq.~\eqref{eq:error_phy} is markedly reduced (by about 50$\%$).}
\label{fig:loss}
\end{figure}

A key to the success of a machine-learning architecture is the training loss.
A commonly used loss function is the mean square error (MSE):
\begin{equation} \label{eq:error_MSE}
\mathcal{L}_\text{MSE}=\| Q_\text{pred}-Q_\text{true} \|^2,
\end{equation}
which typically works well when the data values are within the same order of
magnitude. For the multilayer perceptron in Fig.~\ref{fig:schematic}(c),
the weights and biases can be solved by the optimization algorithm based
on the loss function. The structure from the input vector $\bm{V}$ to the
output vector $\bm{Q}$ is mathematically designed without taking
into account physical constraints of the underlying quantum scattering system.
As a result, even with extensive training, nonphysical results can arise,
which are in fact not uncommon. For example, to realize cloaking requires that
the scattering efficiency have near zero values, but some components of the
output vector $\bm{Q}$ can be negative, which is not physical, as illustrated
in Fig.~\ref{fig:loss}(a) and the inset for a graphene quantum dot that has
$N = 9$ layers. This problem cannot be fixed through training. It is thus
necessary to take physical constraints into account in the design of the
neural-network architecture.

Incorporating physical principles and constraints into designing neural
networks has become a recent subarea of research in machine learning. For
example, the Hamiltonian structure has been built into the neural-network
architectures to predict the dynamical behaviors of classical mechanical
systems~\cite{de1993class,greydanus2019hamiltonian,toth2019hamiltonian,bertalan2019learning,choudhary2019physics,chen2019symplectic,cranmer2020lagrangian,choudhary2021forecasting,han2021adaptable}. Physics-based neural networks have also
been extended to other fields such as fluid
dynamics~\cite{xiong2020neural,gao2021super} and meta surface design in
optical~\cite{jiang2019global} or quantum
systems~\cite{sehanobish2020learning,han2021tomography}. For our quantum
scattering system, a basic requirement is that the scattering cross sections
or efficiency not be negative. We address the issue of physical constraints by
incorporating an additional output layer - the rightmost layer in
Fig.~\ref{fig:schematic}(c) with the output vector
$\bm{\hat{Q}} \equiv (\hat{q}_1,\hat{q}_2,\ldots,\hat{q}_M)^T$. More
specifically, the additional output layer takes the vector $\bm{Q}$ as input
and performs an exponential operation on each component of $\bm{Q}$ to
generate the vector $\bm{\hat{Q}}$. Mathematically, the operation can be
represented as $\mathcal{R}^{-1}(\cdot)\equiv \exp (\cdot)$, which maps $Q$ to
a strictly positive function $\hat{Q}$. The consideration has led us to
introduce the following two-component loss function:
\begin{equation} \label{eq:error_phy}
\mathcal{L}_\text{Physical Loss}=\| Q_\text{pred}-\mathcal{R}(Q_\text{true}) \|^2 + \| \hat{Q}_\text{pred}-Q_\text{true}\|^2,
\end{equation}
where $Q_\text{pred}$ is the scattering efficiency calculated from the vector
$\bm{Q}$ and $\hat{Q}_\text{pred}=\mathcal{R}^{-1}(Q_\text{pred})$. Note that
the squared difference between the true $Q$ function and the transformed
function $\hat{Q}$ is the conventional loss function. The idea to enforce the
physical rule of the nonnegativeness of the scattering cross section is to
supply an additional term: the squared difference between the transformed
true output function $\mathcal{R}(Q_{\rm true})$ and the original output
function $Q$. From the point of view of optimization, this loss function works
as follows. If $Q_\text{true}$ is close to zero, the operation $\mathcal{R}$
will return a relatively large value. When $Q_\text{true}$ is large, the second
term in the loss function will dominate. For small and large values of
$Q_\text{true}$, the predicted values of the scattering efficiency will be
positive.

The performance of the physics-constrained loss function is illustrated in
Fig.~\ref{fig:loss}(b), where various values of the predicted scattering
efficiency function is plotted against the corresponding true values. Compared
with Fig.~\ref{fig:loss}(a) based on the mathematical loss function in
Eq.~\eqref{eq:error_MSE}, we see that the predicted values are all positive
(as they should be) and are closer to the true values. The prediction
performance can be characterized by the following relative error:
\begin{equation} \label{eq:Relative_Error}
\text{Relative Error} =\left| \frac{Q_\text{true}-Q_\text{pred}}{Q_\text{true}} \right|.
\end{equation}
Figure~\ref{fig:loss}(c) shows the error versus the number of layers in the
graphene quantum-dot structure, where each data point is generated from an
ensemble average of $100$ independent neural-network realizations and, for each
realization, the testing dataset contains 1000 points. Two sets of results
are shown: one according to the physics-constrained loss function
Eq.~\eqref{eq:error_phy} (blue points) and another based on the loss function
Eq.~\eqref{eq:error_MSE} (red). We see that, as the number of layers in the
quantum dot increases, the architecture becomes more complex, leading to some
increase in the relative error. However, the relative error can be markedly
reduced by employing the physics based loss function Eq.~\eqref{eq:error_phy}.

\subsection{Principle of inverse design} \label{subsec:PID}

\begin{figure*} [ht!]
\centering
\includegraphics[width=\linewidth]{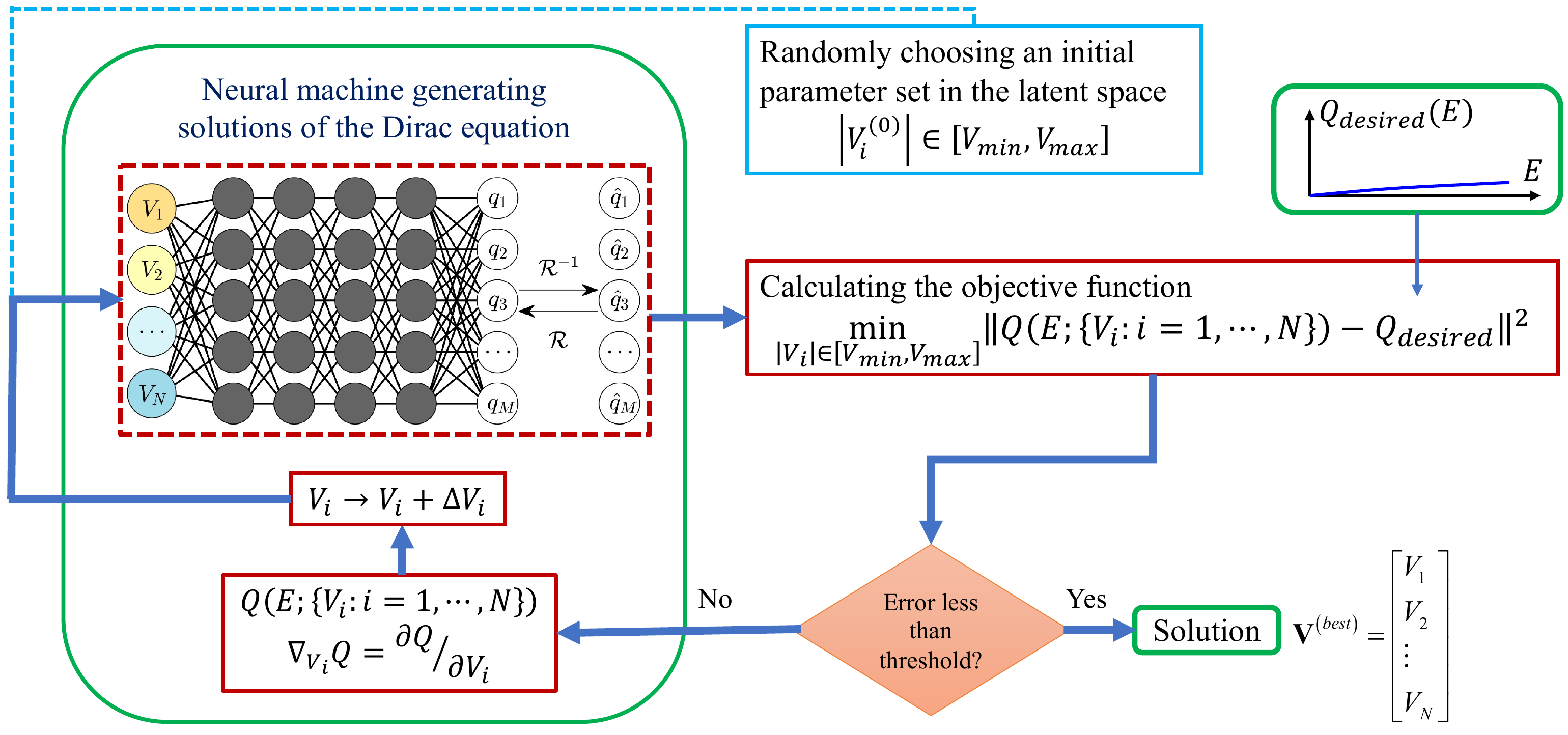}
\caption{Flow chart of machine-learning based solution to the inverse-design
problem of quantum scattering. A multilayer neural network is first trained
using a number of functions $Q(E)$ of the scattering efficiency versus the
electron energy for scattering from a multilayer graphene quantum dot subject
to externally applied gate potentials, one to each graphene layer. For any
given set of gate potentials, the corresponding function of the scattering
efficiency is obtained from the exact solutions of the Dirac equation. A well
trained neural network will generate solutions of the Dirac equation. Solving
the inverse-design problem requires the following steps: (1) choosing a random
initial set of gate potentials as the input to the neural network to obtain
the $Q$ function, (2) calculating the difference between this and the desired
$Q$ function, (3) if the difference is sufficiently small (e.g., less than a
pre-defined threshold), then a solution is deemed to be found, (4) otherwise
calculating a change in each gate potential through the standard gradient
approach applied to the neural network to yield a new set of input variables,
and (5) repeat steps (3) and (4) until a solution is found.}
\label{fig:flow}
\end{figure*}

The physics-constrained neural network as shown in Fig.~\ref{fig:schematic}(c),
once trained, takes an input vector $\bm{V}$ of the gate potentials to generate
an output vector $\hat{\bm{Q}}$ representing the scattering efficiency that is
the solution of the 2D Dirac equation. A well trained neural network thus
effectively functions as a Dirac equation solver. The main advantage of this
substitution lies in the computational efficiency, as what it takes for the well
trained neural network to generate the solutions of the Dirac equation is
simply matrix multiplications through a gradient descent. More specifically, let
$\bm{Q}_\text{desired}$ be the desired function of the scattering efficiency in
the same energy range as the training datasets, represented as a vector. The
objective is to solve the following optimization problem:
\begin{equation} \label{eq:inverse}
	\underset{\mathbf{V}}{\min}  \text{ }\left\| \frac{\hat{\bm Q}(\mathbf{V})- {\bm Q}_\text{desired}}{{\bm Q}_\text{desired}}  \right\|^2,
\end{equation}
where $\hat{\bm Q}$ is the output vector from the neural network and
$\bm{Q}_\text{desired}$ is normalized to accommodate the behavior of the
scattering efficiency on multiple energy scales. The error in
(\ref{eq:inverse}) to be minimized measures the functional distance between
neural-network generated scattering efficiency and the desired one. In general,
there is no guarantee that a particular scattering configuration can be found
to yield the desired function $\bm{Q}_\text{desired}$, so the error can be
large. The goal is to find the optimal input vector $\bm{V}$ to minimize the
error in (\ref{eq:inverse}). As we will describe below, this can be done in
a computationally efficient manner. Note that the error in (\ref{eq:inverse})
is different from the training error in Eq.~\eqref{eq:Relative_Error}, which
measures the difference between the neural network output to the ground truth,
i.e., the scattering efficiency function from a given set of input gate
potentials.

To better describe the solutions of the optimization problem, we employ a
concrete parameter setting. To generate either cloaking or superscattering,
we use a small or a large quantum dot, respectively, as a reference to obtain
the desired function ${\bm Q}_\text{desired}$, subject to the average gate
potential $87.5$ meV. Initially, the various gate potentials are chosen
randomly from the two symmetric intervals defined by $|V_i|\in [75, 100]$ meV.
Let the desired energy region be from $25$ meV to $70$ meV, which is
discretized into $200$ points, on which both the $\bm{Q}_\text{desired}$ and
$\hat{\bm Q}(\mathbf{V})$ functions are evaluated. The number $N$ of layers
in the graphene quantum dot depends on the design imperative, i.e., whether
cloaking or superscattering is to be realized. Given a task, once $N$ has been
determined, it is fixed so that training and inverse design are carried out on
the same neural-network architecture. After training, the error function in
\eqref{eq:inverse} is a deterministic function, whose global minimum can be
found by using, e.g., the standard interior point
method~\cite{byrd1999interior,byrd2000trust}. We incorporate a gradient descent
procedure into the interior point method as it was established previously that
this can significantly improve the computation
speed~\cite{peurifoy2018nanophotonic}.

A detailed description of the steps involved in finding an optimal solution
to the inverse-design problem is presented in Fig.~\ref{fig:flow}.

Four remarks are in order. First, using the trained neural network in
Fig.~\ref{fig:flow} as a Dirac-equation solver has the advantage of being
extremely computational efficient. If the original Dirac equation were used
for calculating the scattering-efficiency $Q$ function and for finding the
optimal set of gate potentials, a vast number of combinations of the potentials
would typically be needed and, for each combination, the calculation of the
$Q$ involves calculating a large number of sophisticated mathematical functions.
With the neural network, the standard gradient-descent method can be invoked
to find the changes in the gate potentials, quickly yielding the optimal
solution. For each given set of input variables, all needed for the neural
network is matrix multiplications, which can be done extremely efficiently
using well-developed packages.

Second, compared with designing a multilayer
spherical scattering structure in optics where the input vector is chosen
from a single interval of the physical parameter (e.g., dielectric constant),
for graphene the input gate potentials are chosen from two symmetric intervals:
one negative and another positive. There is then a gap in the input parameter
space. To our knowledge, this type of inverse design has not been studied. Our
solution is to test all possible combinations for
$V_i\in [-100, -75]\cup [75, 100]$ meV. For each $V_i$, there are two
possibilities: positive or negative. For an $N$-layer circular quantum dot,
altogether we have $2^N$ possible combinations. Take $N=2$ as an example.
If we wish to find the best $V_{(1,2)}$ such that $|V_{(1,2)}|\in [75,100]$
meV, there are four combinations:
$V_1\in[75, 100]$ meV $\cap$ $V_2\in[75, 100]$ meV,
$V_1\in[75, 100]$ meV $\cap$ $V_2\in[-100, -75]$ meV,
$V_1\in[-100, -75]$ meV $\cap$ $V_2\in[75, 100]$ meV, and
$V_1\in[-100, -75]$ meV $\cap$ $V_2\in[-100, -75]$ meV.  For each combination,
an optimal solution can be found using one or two optimization steps. We find
that, for a larger value of $N$, the best scattering structure can be found
with one or two optimization steps for each combination of the gate potentials.
As a result, for any reasonable value of $N$ (e.g., $N = 10$), the required
computational complexity is well manageable.

Third, to find the global minimum in \eqref{eq:inverse}, including derivatives
can increase the computational efficiency~\cite{peurifoy2018nanophotonic}. In
particular, to optimize $\mathbf{V}$ from \eqref{eq:inverse}, we calculate the
derivative $\partial \text{Loss}/\partial \mathbf{V}$. Since $Q_\text{desired}$
is fixed, it is only necessary to evaluate the derivative of
$\hat{Q}(\mathbf{V})$ with respect to $\mathbf{V}$. Once the neural network has
been trained, such derivative can be evaluated through back propagation.

Fourth, there is no guarantee that the ``best'' scattering configuration
can be found to make the scattering efficiency function match the desired
function. Empirically, we find that increasing the number $N$ of graphene
layers typically results in a smaller optimization loss. This issue will be
addressed in detail in Sec.~\ref{sec:issues}.

\section{Results} \label{sec:result}

\subsection{Cloaking} \label{subsec:cloaking}

The goal of designing a graphene-based cloaking system is to generate near-zero
scattering cross section or efficiency from a relatively large quantum dot
structure. Since a small structure tends to generate minuscule scattering
efficiency, the desired vector ${\bm Q}_\text{desired}$ can be found by solving
the relativistic quantum scattering problem over a small single-layer structure,
e.g., a circular cavity of radius $R_C=3$ nm subject to a gate potential of
$V=87.5$ meV, which produces near-zero scattering efficiency, as shown in
Fig.~\ref{fig:schematic}(b). Now consider a nine-layer ($N = 9$) structure of
size ten times larger: $R_N =30$ nm. The question can be stated as: is it
possible to find a set of suitable gate potentials represented by input vector
${\bm V}$ such that the resulting scattering efficiency function is arbitrarily
close to ${\bm Q}_\text{desired}$, subject to the constraint
$\sum|V_i|/N\approx 87.5$ meV?

\begin{figure} [ht!]
\centering
\includegraphics[width=\linewidth]{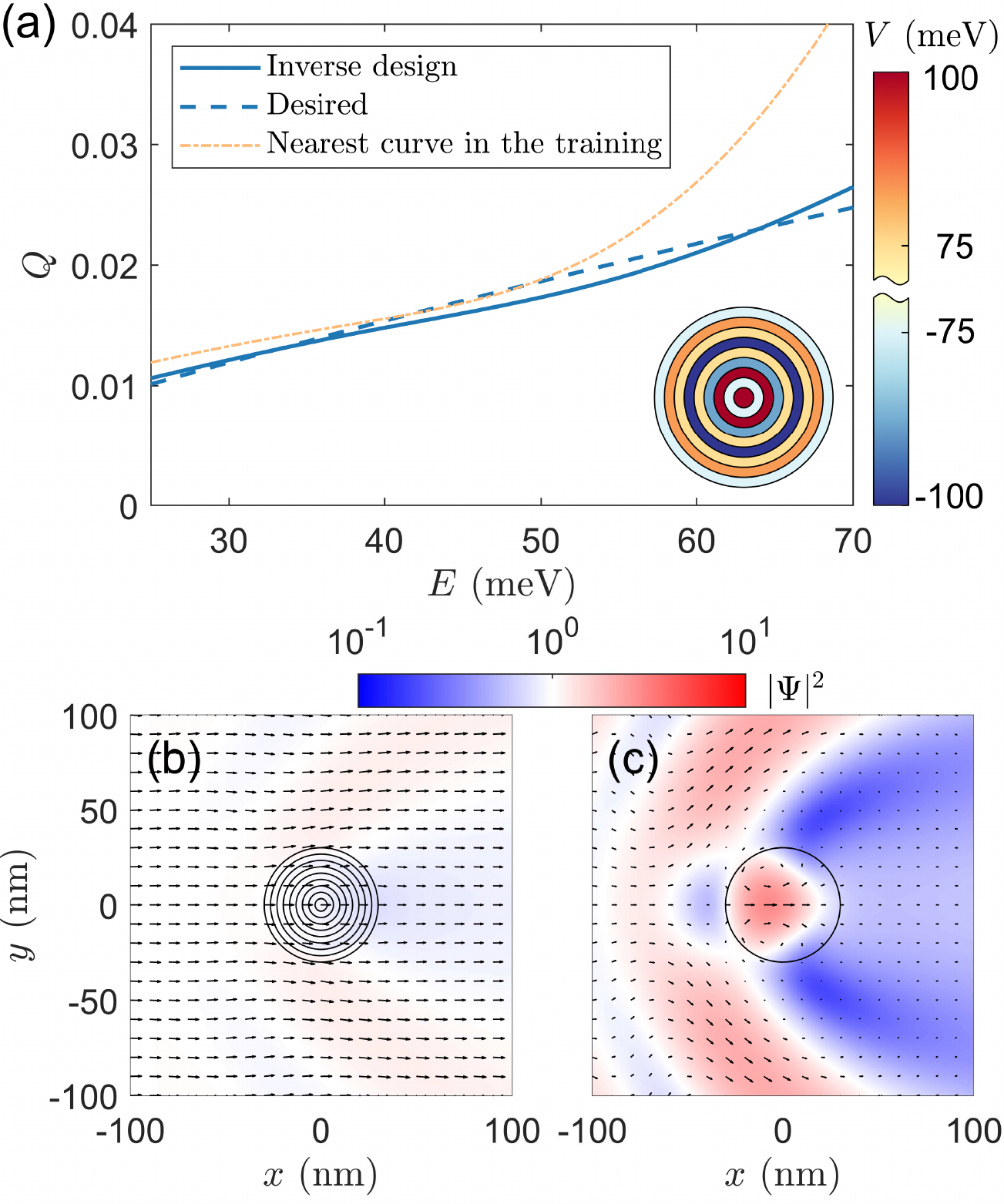}
\caption{Optimal multilayer graphene structure exhibiting cloaking found
by machine learning. The parameter requirements of inverse design are
$R_N = 30$ nm and $\sum|V_i|/N\approx 87.5$ meV. (a) Comparison of three curves
of scattering efficiency: the desired function from a single-layer scatterer
of ten times smaller (dashed blue curve), the function underlying cloaking
from the neural network (solid blue curve), and a curve closest to the target
curve from the training dataset (dotted orange curve). The inset illustrates
the optimal quantum-dot structure predicted by machine learning. That the two
blue curves stay close to each other in the entire energy range is indicative
of cloaking, due to the much larger size of the neural-network produced
multilayer structure. (b) Spatial patterns of the spinor wavefunction and the
local currents associated with cloaking, with the quantum-dot structure at the
center and an incident plane wave from the left. The wavefunction intensity is
nearly constant (unity) in the relevant spatial domain and the current vectors
hardly change their directions, signifying cloaking. (c) Conventional
scattering patterns from a single-layer quantum dot of the same size. There is
no cloaking due to the significant changes in both the wavefunction intensity
and the current direction across the domain.}
\label{fig:cloaking}
\end{figure}

Figure~\ref{fig:cloaking}(a) presents three curves of the resulting scattering
efficiency function: the desired function (dashed blue), the function from the
optimal multilayer graphene structure found by our physics-constrained neural
network (solid blue curve), and a training curve that is the nearest to the
target curve from the training dataset (dotted orange curve). The optimal
physical structure of the multilayer graphene quantum dot found by our
physics-constrained neural network is shown in the lower right inset, where
the gate potential values applied to different layers are specified by the
color bar with $\sum|V_i|/N=85.7$ meV. We see that the optimal structure
produces a scattering efficiency function that well approximates the desired
function. The purpose of the dotted orange curve is to demonstrate that the
optimal structure is not a trivial interpolation of some of structures in the
training dataset. Apparently, this ``best'' training function deviates from the
desired function as the energy increases. The structure found by the neural
network is thus one that produces real cloaking in that the scattering
efficiency is as small as that produced by a structure of ten times smaller in
the entire energy range of interest.

Intuitively, the emergence of cloaking from a large scattering structure
requires destructive quantum interference. If the gate potentials applied to
nearby layers have the same sign, it will be difficult to produce such
interference. However, alternating signs of the gate potentials between
neighboring layers can facilitate the occurrence of destructive interference.
The pattern of gate potentials of the optimal multilayer structure predicted
by the neural network indeed has this feature, as demonstrated in
Fig.~\ref{fig:cloaking}(a).

To visualize the cloaking phenomenon, we show the spatial distribution of the
scattering wavefunction. In particular, we compare two cases: the optimal
nine-layer structure found by the neural network and a single-layer quantum dot
of the same size ($R = 30$ nm), in terms of the wavefunction intensity defined
as $|\psi_1|^2+|\psi_2|^2$, where $\psi_1$ and $\psi_2$ are the two spinor
components and the directions of the local current vectors are given by
$[\langle\sigma_x\rangle, \langle \sigma_y\rangle]$. For a planar incident wave
of unit intensity from the left, if cloaking has indeed occurred, the
wavefunction intensity should be nearly constant in the spatial domain of
interest and the current vectors should not change their directions, i.e., to
maintain their horizontal directions. Such behaviors characteristic of cloaking
are exemplified in Fig.~\ref{fig:cloaking}(b) for the optimal multilayer
structure. In contrast, for the single-layer structure where no cloaking
occurs, the wavefunction intensity varies across the domain and there are
significant changes in the directions of the local current vectors, as shown
in Fig.~\ref{fig:cloaking}(c).

For a graphene quantum-dot scatter with a weak scattering strength as measured
by $\bar{V}R$, where $\bar{V}$ is the average potential and $R$ is the dot
size, the scattering efficiency with the strength exhibits the following
scaling relation~\cite{wang2021super}: $Q \sim (\bar{V}R)^3$. The target quantum
dot used for generating the cloaking behavior has the size of 3 nm. According
to this scaling law, the scattering efficiency should be quite small. For a
relatively large circular structure of size, e.g., $10$ nm, the scattering
efficiency saturates and the scaling law no longer holds, but it can still
be used to obtain an order-of-magnitude estimate of the scattering efficiency.
We find that for a conventional structure of size $30$ nm without cloaking, the
scattering efficiency is typically two orders of magnitude larger than that
from a structure of size ten times smaller. It is remarkable that our
physics-constrained machine learning approach is able to predict a large
quantum-dot structure but with scattering efficiency two orders of magnitude
smaller than it ``should be'' in the conventional sense, effectively realizing
cloaking.

\subsection{Superscattering} \label{subsec:superscattering}

\begin{figure} [ht!]
\centering
\includegraphics[width=\linewidth]{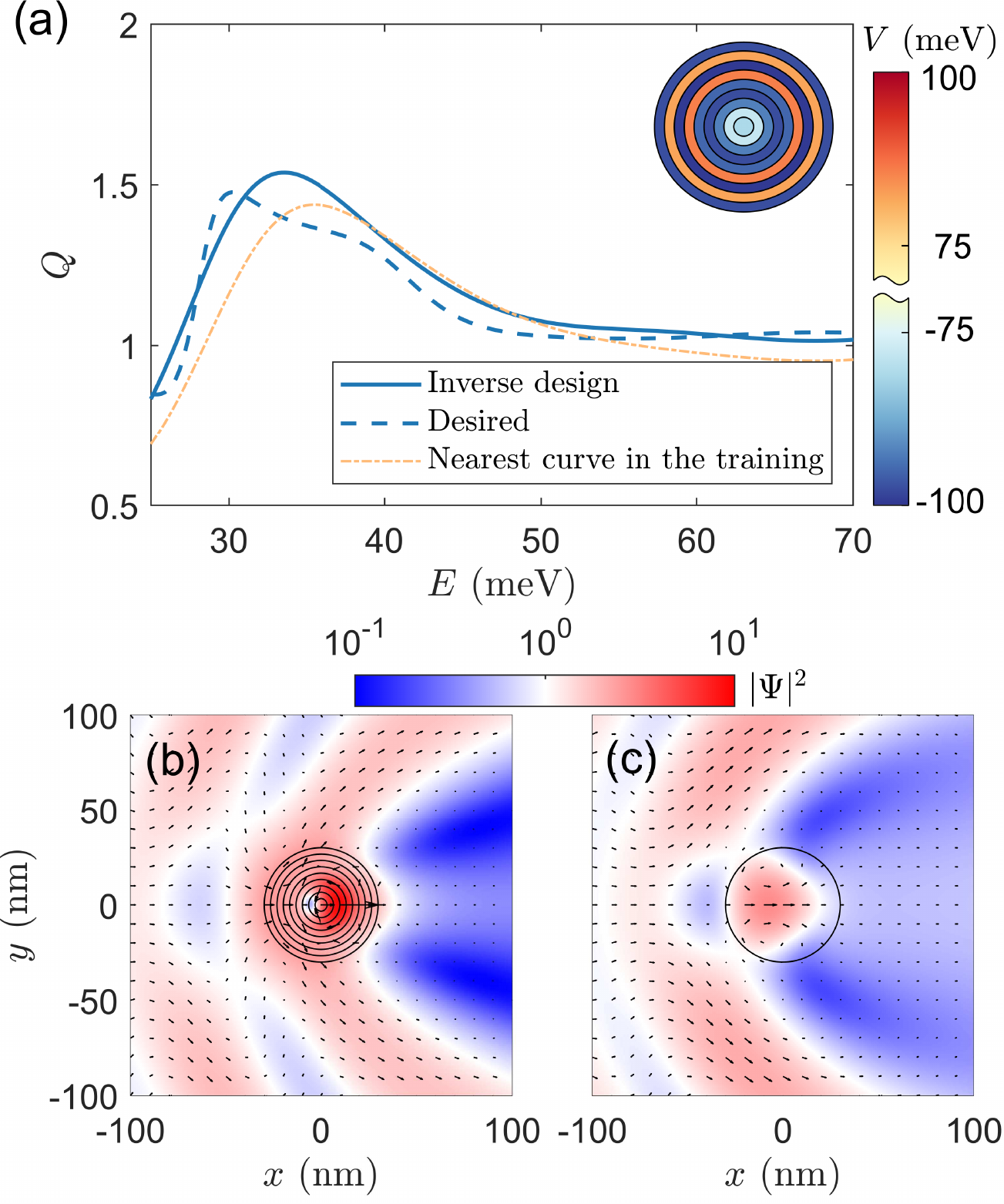}
\caption{Optimal multilayer graphene structure exhibiting superscattering
produced by machine learning. As for the case of cloaking in
Fig.~\ref{fig:cloaking}, the structure has the size $R_N = 30$ nm and the gate
potentials satisfy the constraint $\sum|V_i|/N\approx 87.5$ meV. (a) Comparison
of three curves of scattering efficiency: the desired function from a
single-layer scatterer of twice larger size (dashed blue curve), the function
associated with superscattering from machine learning (solid blue curve), and a
training curve closest to the target curve from the training dataset (dotted
orange curve). The upper right inset illustrates the optimal quantum-dot
structure predicted by the neural network. (b) Spatial patterns of the spinor
wavefunction and the local currents associated with superscattering at the
energy $E=30$ meV, with the quantum-dot structure at the center and an incident
plane wave from the left. There are significant variations in the wavefunction
intensity and in the directions of the local current vectors across the spatial
domain of interest. (c) Conventional scattering patterns from a single-layer
quantum dot of the same size.}
\label{fig:SS}
\end{figure}

We now address the opposite (to cloaking) problem: is it possible to design
a multilayer scattering structure of certain size to generate scattering
efficiency that conventionally would be achieved only with a much larger
scatterer? To be concrete, we use a nine-layer structure of size $R_N = 30$ nm
with gate potentials subject to $\sum|V_i|/N\approx 87.5$ meV and generate the
target scattering efficiency function ${\bm Q}_\text{desired}$ using a
single-layer scatterer of size $R_A=63$ nm, as illustrated in
Fig.~\ref{fig:schematic}(b). The actual curve of ${\bm Q}_\text{desired}$ is
shown by the dashed blue curve in Fig.~\ref{fig:SS}(a), where a scattering
resonance arises at energy about $E=30$ meV. Applying our machine-learning
based inverse design algorithm, we obtain the quantum-dot structure as
illustrated in the upper right corner of Fig.~\ref{fig:SS}(a) with
$\sum|V_i|/N=89.8$ meV, where the resulting scattering efficiency function is
represented by the solid blue curve that stays near the target curve in the
entire energy range. This means that the optimal structure of size $30$ nm
is able to produce scattering efficiency as large as that from a structure of
much larger size. For reference, the ``best'' scattering efficiency function
from the training dataset is shown (the dotted orange curve). The optimal
quantum-dot structure discovered by the neural network for superscattering
requires a negative gate potential for majority of the layers. The spatial
distribution of the wavefunction intensity and the local current vectors
associated with superscattering are shown in Fig.~\ref{fig:SS}(b), and
the same quantities from a single-layer scatterer of the same size is shown
in Fig.~\ref{fig:SS}(c). It can be seen that scattering from the optimal
multilayer structure is markedly enhanced. An advantage of our machine-learning
design is that no additional training of the neural network is required here,
insofar as it has been trained for cloaking with a predefined set of scattering
configurations, each with a distinct scattering efficiency curve. That is, the
same neural network used to realize cloaking can be used to find a system that
exhibits superscattering.

Comparing Fig.~\ref{fig:SS} with Fig.~\ref{fig:cloaking}, we see that the
effect of scattering enhancement is not as pronounced as that associated with
scattering suppression in cloaking. The physical reason is Klein tunneling.
In particular, the median Fermi energy is about half of the maximum gate
potential: $E\approx |V|/2$, for which Klein tunneling is significant in
graphene~\cite{KNG:2006}. As a result, a spinor wave tends to change little
in its wavevector after entering the scattering region. To generate strong
cloaking in graphene is thus much more likely than to produce superscattering.
We note that Klein tunneling is a unique feature of Dirac waves.

\section{Issues pertinent to inverse design} \label{sec:issues}

\subsection{Effect of the number of quantum-dot layers} \label{subsec:N}

Physically, the objective of machine-learning based inverse design is to find
a multilayer quantum dot structure to generate cloaking or superscattering by
tuning a set of experimentally feasible, external parameters. From a
mathematical point of view, the task is to find a complicated multilayer
scattering structure to achieve a desired function. While increasing
the number of layers in general allows more complicated scattering functions,
the computational complexity increases as well. From the point of view of
machine learning, a more sophisticated scattering structure requires more
training data.

\begin{figure} [ht!]
\centering
\includegraphics[width=\linewidth]{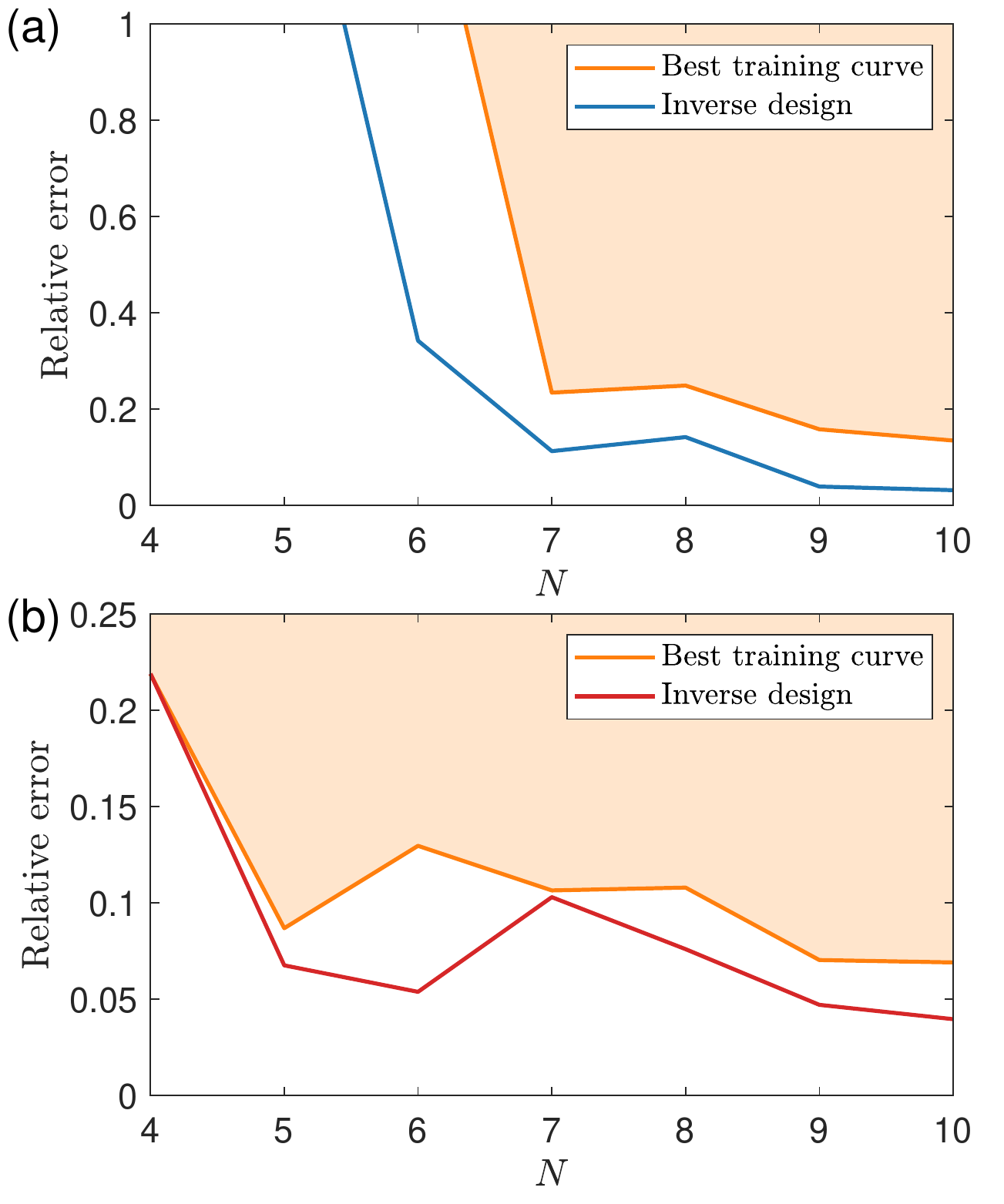}
\caption{Effect of the number of quantum-dot layers on training and
inverse-design errors. (a) For designing a cloaking device, the errors versus
$N$, where the shaded region indicates the normalized mean square error from
the training dataset and the solid blue curve represents the error between
the scattering function from the optimal structure and the desired function.
As the number of quantum-dot layers increases, the errors decrease. (b) Similar
behaviors arise when designing a superscattering device.}
\label{fig:N}
\end{figure}

Figure~\ref{fig:N}(a) shows, when designing a cloaking device, the relative
training and inverse-design errors versus $N$, where the shaded area denotes
all possible error values from training, with the solid orange curve denoting
the minimally possible training error. Physically, a circular quantum dot of
too few layers is unable to lead to cloaking, so the errors are large for small
$N$ values. As $N$ increases, it is more likely to generate destructive
interference to realize cloaking, leading to a decrease in the errors. The
solid blue curve in Fig.~\ref{fig:N}(a) represents the error in the scattering
function from the optimal dot structure found by inverse design, which is below
the minimum training-error curve. That is, the neural network is able to find a
quantum-dot structure whose scattering function is closer to the desired one
than any function in the training dataset. A similar behavior arises for the
task of generating superscattering, as shown in Fig.~\ref{fig:N}(b).

The results in Fig.~\ref{fig:N} indicate that there is no guarantee that an 
increase in the number of layers can lead to better inverse design. In our 
work, the set of radii of the graphene quantum dot is fixed to be 
$R_i=i R_N/N$. Suppose $N$ is increased by some integer factor. The resulting 
quantum dot would lead to a better chance of realizing scattering possibilities
but the structure becomes more sophisticated. In general, the performance of 
the new structure should be improved, but empirically this is true only when 
the numbers of layers of the two structures are commensurate. From another 
angle, a perturbation on the gate potential of a graphene layer will affect 
the resulting error in the scattering efficiency but, as will be shown below 
in Fig.~\ref{fig:RO}, and modifying the gate potential of an outer layer has 
a stronger effect on augmenting the error than disturbing the potential of an 
inner layer. This suggests that the non-monotonous behavior in the error be 
an intrinsic physical property of the scattering system that has little 
dependence on the machine learning procedure.

Due to the physical nature of the error behavior, choosing the ``best'' number
of graphene layers is difficult. In fact, the same problem arises in
optical inverse-design problems. For example, in 
Refs.~\cite{peurifoy2018nanophotonic,sheverdin2020photonic}, 
it was necessary to pre-assign the number of neural-network layers. A 
mathematical difficulty is that this number is not a continuous variable. In 
Refs.~\cite{sajedian2019optimisation,so2020deep}, reinforcement learning
was used to find the best device whose dimension is only allowed to take on
some discrete value. Another difficulty is that the multilayer perceptron
employed in our work is suitable for inputs with a fixed dimension. In
the field of computer vision, a method called Spatial Pyramid 
Pooling~\cite{he2015spatial} was proposed to address this difficulty. The idea 
was extended to adaptive pooling layers in PyTorch~\cite{paszke2019pytorch}. 
While for our quantum-dot inverse-design problem, increasing the number of 
layers can lead to better performance in most cases, to the best of our 
knowledge to predetermine the optimal number of layers before training remains
to be an open problem.

\subsection{Robustness of inverse design} \label{subsec:EID}

\begin{figure} [ht!]
\centering
\includegraphics[width=\linewidth]{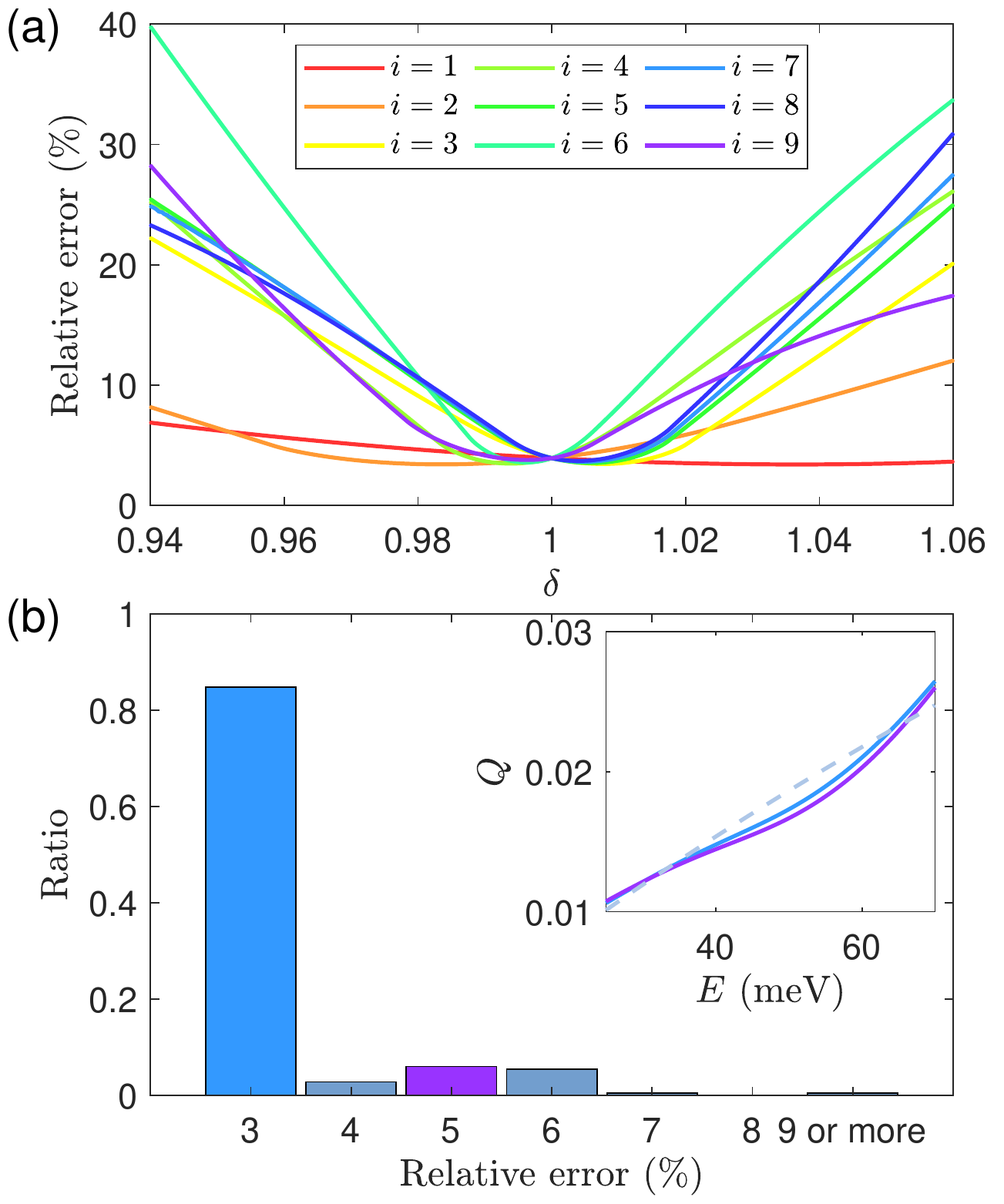}
\caption{Accuracy and robustness of inverse design. (a) For a graphene quantum
dot of $N = 9$ layers to realize cloaking, normalized mean square error in the
machine generated scattering efficiency function versus the perturbation
parameter $\delta$ in the gate potential as applied to different layers of the
graphene quantum dot, where $\delta = 1$ corresponds to zero perturbation. The
error is calculated from the exact solution of the Dirac equation. The ideal
global minimum of the loss function should occur at $\delta = 1$. The actual
minima are quite close to $\delta = 1$. (b) Histogram of the relative error in
the machine-generated scattering efficiency function from an ensemble of $10^4$
combinations of the input parameters. More than $80\%$ of times, the global
minima are within $3\%$ of $\delta = 1$, indicating robustness of the
inverse-design scheme. The inset shows two examples of the scattering
efficiency function from the machine-predicted optimal dot structure, with the
relative error of $3\%$ and $5\%$, respectively, where the desired function
is indicated by the dashed curve.}
\label{fig:RO}
\end{figure}

The main idea underlying machine-learning based inverse design is that a
properly trained neural network is able to generate an approximation of a
desired mathematical function. In general, the optimal physical structure
found by machine learning is close to but not necessarily the global minimum
of the loss function Eq.~\eqref{eq:inverse}. It is useful to examine how
``far away'' the loss associated with the optimal structure is from the global
minimum. Consider the ``best'' quantum-dot structures for cloaking and
superscattering design, as shown in Figs.~\ref{fig:cloaking} and \ref{fig:SS},
respectively, where the structure contains nine layers so the loss function
depends on nine gate-potential variables and is thus nine-dimensional. Suppose
the optimal structure is described by the following set of $N$ potential
values: $[V_1,V_2,\cdots,V_N]$. We apply a small perturbation e.g., $V_1$,
which leads to a new set of potentials: $[V_1\delta,V_2,\cdots,V_N]$, and
examine the dependence of the loss function on $\delta$. If the original
solution $[V_1,V_2,\cdots,V_N]$ corresponds to the global minimum, the loss
function versus $\delta$ should exhibit a minimum at $\delta = 1$.
Figure~\ref{fig:RO}(a) shows the loss $\mathcal{L}^{(i)}$ versus $\delta$ for
different layer index $i$. The functions are approximately parabolic with
its minimum close to $\delta = 1$ but not exactly at $\delta = 1$, because
both the neural network and the optimization algorithm are approximate solvers
and the resulting scattering structure is not the absolute best. When the
potential change occurs on one of the innermost layers (corresponding to small
layer indices), the parabolic functions are rather shallow, but they are
steep for large layer indices, indicating that modifying the gate potential
of an outer layer has a stronger effect on augmenting the error. A plausible
reason is that an outer layer has a larger area than that of an inner layer
for a fixed layer width.

As discussed in Sec.~\ref{subsec:PID}, in comparison with the inverse problem
of electromagnetic wave scattering, a difficulty in solving the inverse problem
of quantum scattering in graphene is that the input parameters can be selected
from a positive and a negative interval, leading to a gap in the between the
two allowed parameter intervals. Implementing an optimization algorithm, e.g.,
the interior point method, requires examining a large number of parameter
combinations from the two intervals. For an $N$-layer graphene quantum dot,
the optimization needs to be done $2^N$ times, requiring that the algorithm be
computationally efficient. In general, even based on a fixed nonlinear function,
the optimization result would depend on the initial condition.
Figure~\ref{fig:RO}(b) shows, for $N=9$ and cloaking design, the statistical
distribution of the relative error of the optimization algorithm from $10^4$
different initial conditions. It can be seen that for more than $80\%$ of the
initial conditions, the optimization errors are about $3\%$, indicating the
robustness of our machine-learning based inverse-design scheme.

\subsection{Computational time required for training and convergence} \label{subsec:Time}

\begin{figure} [ht!]
\centering
\includegraphics[width=\linewidth]{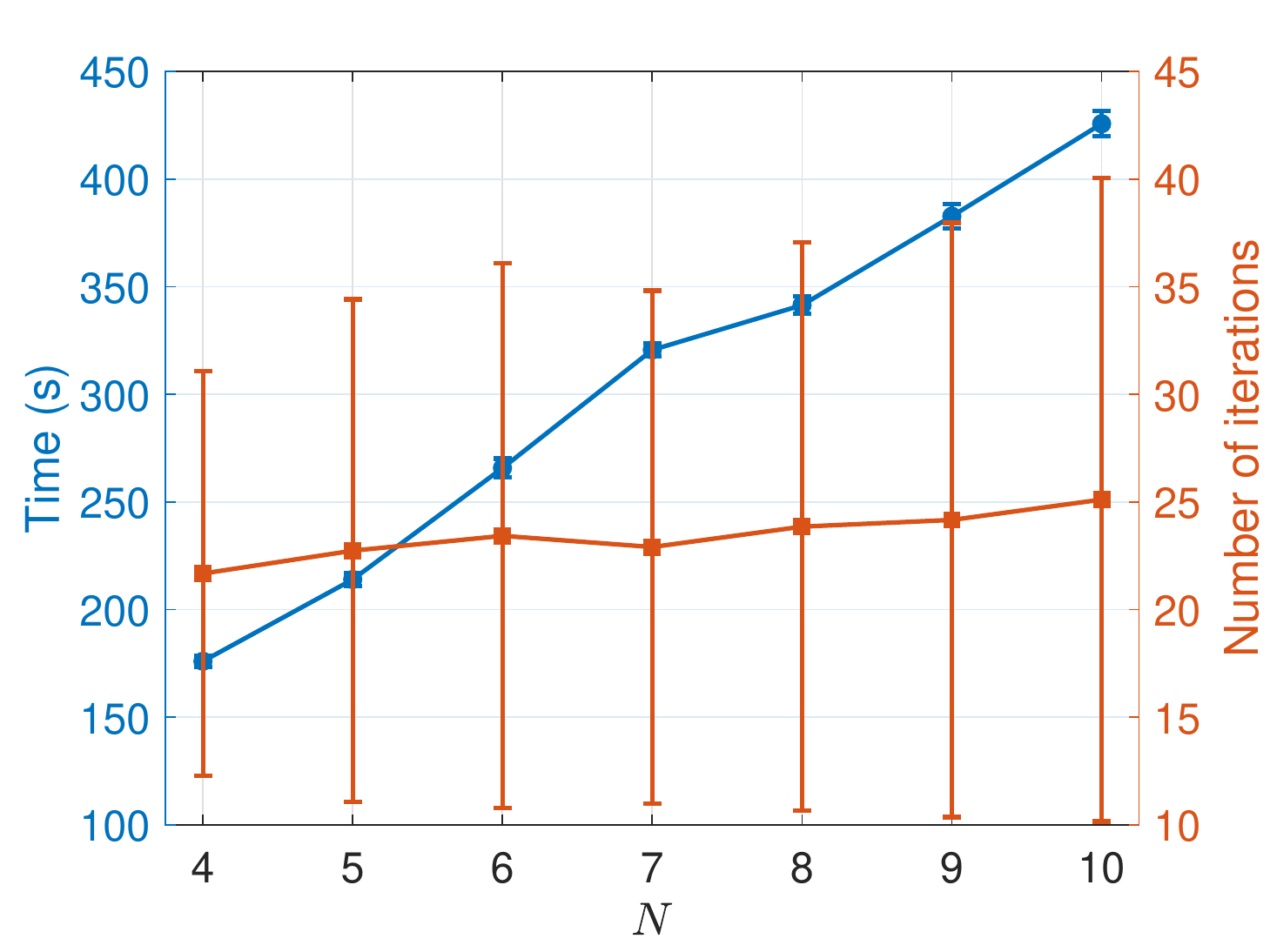}
\caption{Computational time required for training and convergence. The data
points associated with the left and right $y$-axis represent, respectively,
the average time (with variance) of the training time and the number
of iterations required for the neural network to converge as the number of
layers constituting the quantum-dot structure increases from four to ten.
The training time increases linearly with $N$ but the number of iterations
changes little due to the fact that the number of parameters in the neural
network is approximately invariant as $N$ increases in this range.}
\label{fig:figure_time}
\end{figure}

Two main factors determine the time required for training: the amount
of training data and the number of training parameters. Empirically, the
amount of training data is proportional to the number $N$ of the layers
constituting the quantum-dot structure. With regard to the training parameters,
our neural network contains four hidden layers. For different values of $N$,
the number of neural-network parameters remains approximately constant.
Figure~\ref{fig:figure_time} exemplifies how the average
training time and variance depend on $N$. With CPU i7-6850k without GPU
acceleration, the average training time ranges from $100$ seconds to $400$
seconds as the number of layers of the scattering structure increases from
four to ten.

About the convergence of the training process, we observe that the training
error depends on the initial condition for the neural network and data 
sampling, as demonstrated in Fig.~\ref{fig:loss}(c). However, the variance in 
the error is small, indicating that convergence is always granted for training.

The computational speed for the inverse design depends on two factors: the 
number of iterations required to achieve convergence and the time cost of each
iteration. Because the number of parameters involved in the neural network does
not change significantly with $N$, for each iteration approximately a similar
amount of time is required for calculating the gradient for different values of
$N$. As $N$ increases, while more iterations are needed for optimizing the 
neural network, we find that the required increase in the number of iterations 
is insignificant. For example, as $N$ increases from four to ten, the average 
number of iterations required increases only from about 21 to about 23, as 
shown in Fig.~\ref{fig:figure_time} as well as in Fig.~\ref{fig:RO}(b) where 
more than $80\%$ of the initial conditions lead to satisfactory convergence.

\section{Discussion} \label{sec:discussion}

To develop quantum systems to realize drastically distinct functions under a
different set of externally imposed parameter values is a challenging problem.
For example, for a multilayer graphene quantum-dot system in which each layer
is subject to a gate potential, one may wish to make the system ``invisible''
by generating cloaking with weak scattering under one set of gate potentials,
but as the environment changes the opposite extreme may be desired: strong
scattering (or superscattering defined in a broad sense). Would it be possible
to simply change the set of external gate potentials, while keeping the
physical structure of the system intact, so as to induce a metamorphic
transition in the quantum scattering dynamics from one extreme to the opposite
extreme? In optics, if the goal is to realize one of the two extremes, then
the method of scattering cancellation can be effective~\cite{alu2005achieving,edwards2009experimental,li2016design,qian2019experimental}.
The challenge is whether it is possible to create a system that is capable
of the two extremes of scattering - superscattering and cloaking, in certain
energy range by tuning a set of experimentally adjustable parameters. In
principle, by exploiting a multilayer scattering structure to induce unusually
strong destructive or constructive quantum interference, it should be possible
to meet the challenge. However, the computational complexity associated with a
brute-force search of the parameter space, aided by the exact solution of the
Dirac equation, tends to grow exponentially with the number of layers
constituting the scatterer, making the problem NP hard.

The main idea underlying this work is that machine learning provides an
effective and computationally efficient approach to solving the challenging
inverse-design problem, i.e., to generate the two opposite extremes of quantum
scattering by tuning a set of experimentally accessible parameters. For the
concrete setting of a multiple-ring graphene quantum dot structure, the basic
component of the machine learning architecture is a conventional multilayer,
feed forward neural network that takes a set of gate potentials (one applied to
each ring layer of the quantum dot) as input and generate the scattering cross
section or efficiency versus the Fermi energy (scattering function) as the
output. Straightforwardly training this baseline neural-network architecture
using an adequate number of pre-determined scattering functions obtained from
directly solving the Dirac equation reveals that a non-negligible fraction of
the output functions can be nonphysical in that they contain negative values of
the scattering efficiency. Imposing the physical constraint that the scattering
efficiency must be positive through an additional output layer leads to a class
of physics-constrained neural networks. The main accomplishment of this paper
is a successful demonstration that physics-constrained machine learning can
effectively solve the inverse design problem of generating quantum scattering
dynamics at the opposite extremes in the same system.

The key advantage and appealing feature of our machine-learning approach to the
extreme inverse-design problem can be appreciated using the paradigmatic
multilayer graphene quantum dot structure studied in this paper. Given a
desired scattering function, the starting point is a random set of gate
potentials. It is practically impossible for this initial set of input
parameter values to produce a scattering function closed to the desired one,
so adjustments to the inputs are necessary. Solving the Dirac equation for
random parameter perturbations is computationally infeasible not only because
the solutions of the Dirac equation require evaluating a large number of
sophisticated mathematical functions but, more importantly, because of the
NP-hard nature of the search problem. These fundamental difficulties can be
overcome by exploiting machine learning for the following reasons: (1) all
mathematical operations needed is matrix multiplication that can be done
extremely efficiently and (2) the standard gradient descent can guild the
solution in the input-parameter space to approach the optimal one in only a
few steps. We have demonstrated that, typically, the optimal solution is close
to the global minimum of the physics-constrained loss function.

In our present work, we have studied quantum-dot systems in the weak scattering
regime, where the wavevector inside the cavity has the same order of magnitude
as the wavevector outside the cavity. In this case, the scattering efficiency
does not exhibit sharp resonant peaks, in contrast to the case of strong  
resonances where the wavevector inside the cavity is much larger than the
wavevector outside~\cite{han2020electrical}. Empirically, a change in the 
potential profile requires that the neural network be retrained.

In general, inverse problems in physical sciences are challenging, but to
design a system that can generate physical behaviors of two opposite extremes
is even harder. We emphasize that no feasible conventional methods exist which
can be used to solve this type of {\em extreme inverse problems} due to the
following difficulties: the need to search through a high-dimensional parameter
space, the complex relation between the system structure and functions, and the
uncertainty about the existence of a configuration that can deliver the desired
system functions. Our machine-learning based solution represents a step forward
in this area.

\section*{Acknowledgment}
We thank Dr.~H.-Y. Xu for providing great inputs during the initial stage of
this project. This work was supported by AFOSR under Grant No.~FA9550-21-1-0438.

\appendix

\section{Scattering cross section for circular graphene quantum dot} \label{Appendix_A}

\subsection{Single-layer circular graphene quantum dot}

Consider a single-layer graphene quantum dot of radius $R_1$, subject to gate
potential $V_1$, where the circular boundary separates the whole space into
two regions: inside ($r < R_1$) and outside ($r > R_1$) the dot, denoted
as regions 1 and 2, respectively. The energies inside and outside the dot
are $E-V$ and $E$, respectively, giving the corresponding wavevectors as
$k_1=|E-V|/v_g$ and $k_2=|E|/v_g$. For convenience, we define two sign
quantities: $\tau_1\equiv \text{Sign}(E-V)$ and $\tau_2\equiv \text{Sign}(E)$.

The incident wave is
\begin{equation}
\psi_\text{in}(\mathbf{r})=\frac{1}{\sqrt{2}}\binom{1}{\tau_2}e^{ik_2x}.
\end{equation}
Using the Jacobi-Anger identity
\begin{equation}
e^{iz\cos\theta}\equiv \sum_{l=-\infty}^{\infty} i^l J_l(z) e^{il\theta},
\end{equation}
we expand the incident wave in the polar coordinates as
\begin{equation}
\psi_\text{in}(\mathbf{r})=\sum_{l=-\infty}^{\infty}\frac{i^l}{\sqrt{2}}\binom{J_l(k_2r)}{i\tau_2J_{l+1}(k_2r)}e^{il\theta}.
\end{equation}
Inside the quantum dot, we have
\begin{equation}
\psi_1(\mathbf{r})=\sum_{l=-\infty}^{\infty} A^{(1)}_l \binom{J_l(k_1r)}{i\tau_1 J_{l+1}(k_1r)} e^{il\theta},
\end{equation}
and the wavefunction outside the dot is given by
\begin{equation}
\psi_2(\mathbf{r})=\psi_\text{in}+\sum_{l=-\infty}^{\infty} A^{(2)}_l \binom{H^{(1)}_l(k_2r)}{i\tau_2 H^{(1)}_{l+1}(k_2r)} e^{il\theta}.
\end{equation}
Matching the boundary conditions for each angular momentum $l$, we get
\begin{widetext}
\begin{equation}
\begin{split}
\begin{bmatrix}
J_l(k_1R_1) & -H_l^{(1)}(k_2R_1) \\
\tau_1 J_{l+1}(k_1R_1) & -\tau_2 H_{l+1}^{(1)} (k_2R_1)
\end{bmatrix}
\begin{pmatrix}
A^{(1)}_l \\
A^{(2)}_l
\end{pmatrix}
=\frac{1}{\sqrt{2}}
\begin{pmatrix}
i^lJ_l(k_2R_2) \\
\tau_2 i^l J_{l+1}(k_2R_2)
\end{pmatrix}.
\end{split}
\end{equation}
\end{widetext}

\subsection{Multilayer circular graphene quantum dot}

Consider an $N$-layer circular graphene quantum dot, where all the layers are
concentric and have radii $R_1$, $R_2$, $\cdots$, and $R_N$ with the
respective gate potentials $V_1$, $V_2$, $\cdots$, and $V_N$. Let
$k_i=|E-V_i|/v_g$ and $\tau_i\equiv \text{Sign }(E-V_i)$. In region 2, we have
$k_{N+1}=|E|/v_g$ and $\tau_{N+1}=\text{Sign}(E)$. To represent the
wavefunction in each region, we use $A^{(1)}$ for $r<R_1$ and
\begin{equation}
\psi_1(\mathbf{r})=\sum_{l=-\infty}^{\infty} A^{(1)}_l\binom{J_l(k_1r)}{i\tau_1J_{l+1}(k_1r)} e^{il\theta},
\end{equation}
where $r<R_1$. For $R_i<r<R_{i+1}$ we have
\begin{widetext}
\begin{equation}
\begin{split}
\psi_{i+1}(\mathbf{r})=\sum_{l=-\infty}^{\infty}A^{(i+1)}_l\binom{H_l^{(1)}(k_{i+1}r)}{i\tau_{i+1}H_{l+1}^{(1)}(k_{i+1}r)}e^{il\theta}
+ \sum_{l=-\infty}^{\infty}B^{(i+1)}_l\binom{H_l^{(2)}(k_{i+1}r)}{i\tau_{i+1}H_{l+1}^{(2)}(k_{i+1}r)}e^{il\theta}.
\end{split}
\end{equation}
\end{widetext}
For $r>R_N$ we have
\begin{widetext}
\begin{equation}
\begin{split}
\psi_{N+1}(\mathbf{r}) = \sum_{l=-\infty}^{\infty}\frac{i^l}{\sqrt{2}}\binom{J_l(k_{N+1}r)}{i\tau_{N+1}J_{l+1}(k_{N+1}r)}e^{il\theta}
+\sum_{l=-\infty}^{\infty}A^{(N+1)}_l\binom{H_l^{(1)}(k_{N+1}r)}{i\tau_{N+1}H_{l+1}^{(1)}(k_{N+1}r)}e^{il\theta}.
\end{split}
\end{equation}
\end{widetext}
The unknown coefficients are $A_l^{(1)}$, $A_{l}^{(2)}$, $\cdots$, $A_l^{N}$ and
$A_l^{(N+1)}$ as well as $B_l^{(2)}$, $B_l^{(3)}$, $\cdots$, and $B_l^{(N)}$.
Altogether, there are $2N$ unknown parameters to be determined. We use the $N$
boundary conditions to generate $2N$ equations, as follows.
\begin{widetext}
For the boundary at $r=R_1$, we have
\begin{equation}
\begin{split}
A^{(1)}_lJ_l(k_1R_1)-A^{(2)}_lH_l^{(1)}(k_2R_1)-B^{(2)}_lH_l^{(2)}(k_2R_1)&=0, \\
A^{(1)}_l \tau_1 J_{l+1}(k_1R_1)-A^{(2)}_l \tau_2 H_{l+1}^{(1)}(k_2R_1)-B^{(2)}_l \tau_2 H_{l+1}^{(2)}(k_2R_1)&=0. \\
\end{split}
\end{equation}
For the boundary at $r=R_i$, where $1<i<N$, we have
\begin{equation}
\begin{split}
A^{(i)}_l H_l^{(1)}(k_iR_i)+B^{(i)}_l H_l^{(2)}(k_iR_i)-A^{(i+1)}_lH_l^{(1)}(k_{i+1}R_i) -B^{(i+1)}_l H_l^{(2)} (k_{i+1}R_i) &=0, \\
A^{(i)}_l \tau_i H_{l+1}^{(1)}(k_iR_i)+B^{(i)}_l \tau_i H_{l+1}^{(2)} (k_iR_i)-A^{(i+1)}_l \tau_{i+1} H_{l+1}^{(1)}(k_{i+1}R_i)-B^{(i+1)}_l \tau_{i+1}H_{l+1}^{(2)}(k_{i+1}R_i) &=0.
\end{split}
\end{equation}
For the boundary at $r=R_N$, we have
\begin{equation}
\begin{split}
A^{(N)}_l H_l^{(1)}(k_NR_N) +B^{(N)}_l H_l^{(2)}(k_NR_N)-A^{(N+1)}_l H_l^{(1)}(k_{N+1}R_N) &= \frac{i^l}{\sqrt{2}}J_l(k_{N+1}R_N), \\
A^{(N)}_l\tau_NH_{l+1}^{(1)}(k_NR_N)+B^{(N)}_l \tau_NH_{l+1}^{(2)} (k_NR_N)-A^{(N+1)}_l \tau_{N+1} H_{l+1}^{(1)}(k_{N+1}R_N)&=\frac{i^l}{\sqrt{2}}\tau_{N+1} J_{l+1}(k_{N+1}R_N).
\end{split}
\end{equation}
\end{widetext}
Solving this set of linear equations, we get the coefficients $A^{(N+1)}$. The scattering efficiency is given by
\begin{equation}
Q=\frac{2}{k_{N+1}R_N}\sum_{l=-\infty}^{\infty} \left(A^{(N+1)}_l\right)^2.
\end{equation}
In numerical computations, we truncate the summation at $\max|l|=7$ to
achieve the desired accuracy.


%
\end{document}